
%
\input phyzzx
\tolerance=1000
\sequentialequations
\def\rl{\rightline}

\def\r#1{$\bf#1$}

\def\t1{{\tilde 1}}

\def\F{\widetilde F}

\def\AEF{A.E. Faraggi}
\def\DVN{D. V. Nanopoulos}

\def\NPB#1#2#3{Nucl. Phys. B {\bf#1} (19#2) #3}
\def\PLB#1#2#3{Phys. Lett. B {\bf#1} (19#2) #3}

\def\PRL#1#2#3{Phys. Rev. Lett. {\bf#1} (19#2) #3}
\def\PRT#1#2#3{Phys. Rep. {\bf#1} (19#2) #3}

\def\IJMP#1#2#3{Int. J. Mod. Phys. A {\bf#1} (19#2) #3}

\def\l{\langle}
\def\r{\rangle}

\REF\GSW{M. Green, J. Schwarz and E. Witten,
Superstring Theory, 2 vols., Cambridge
University Press, 1987.}
\REF\EU{\AEF, \PLB{278}{92}{131}.}
\REF\TOP{\AEF, \PLB{274}{92}{47}.}
\REF\SLM{\AEF, \NPB{387}{92}{289}.}
\REF\REVAMP{I. Antoniadis, J. Ellis,
J. Hagelin, and \DVN, \PLB{231}{89}{65}.}
\REF\FNY{\AEF, D.V. Nanopoulos and K. Yuan, \NPB{335}{90}{437}.}
\REF\ALR{I. Antoniadis, G.K. Leontaris and J. Rizos, \PLB{245}{90}{16}.}
\REF\DSW{M. Dine, N. Seiberg and E. Witten, \NPB{289}{87}{585}.}
\REF\NRT{\AEF, WIS--92/48/JUN--PH, Nucl. Phys. B, in press.}
\REF\FM{\AEF, WIS--92/81/OCT--PH.}
\REF\ADS{J.J. Atick, L.J. Dixon and A. Sen, \NPB{292}{87}{109};
         S. Cecotti, S. Ferrara and M. Villasante, \IJMP{2}{87}{1839}.}
\REF\KLN{S. Kalara, J. Lopez and D.V. Nanopoulos,\PLB{245}{91}{421};
\NPB{353}{91}{650}.}
\REF\FH{A. E. Faraggi and E. Halyo, WIS-93/3/JAN-PH, Phys. Lett. B, in press.}
\REF\FFF{I. Antoniadis, C. Bachas, and C. Kounnas, \NPB{289}{87}{87};
I. Antoniadis and C. Bachas, \NPB{298}{88}{586};
H. Kawai, D.C. Lewellen, and S.H.-H. Tye,
Phys. Rev. Lett. {\bf57} (1986) 1832;
Phys. Rev. D {\bf 34} (1986) 3794;
Nucl. Phys. B {\bf 288} (1987) 1;
R. Bluhm, L. Dolan, and P. Goddard,
Nucl. Phys. B {\bf 309} (1988) 330.}
\REF\FHn{A.E. Faraggi and E. Halyo, WIS-93/4/JAN-PH, Phys. Lett. B,
in press.}
\REF\YUKAWA{A.E. Faraggi, WIS--91/83/NOV--PH, Phys. Rev. D, in press.}
\REF\GCU{A.E. Faraggi, Phys. Lett. B {\bf302} (1993) 202.}
\REF\GLT{B. Grzadkowski, M. Lindner and S. Theisen,
\PLB{198}{87}{64}.}
\REF\BKS{A. Bouquet, J. Kaplan and C.A. Savoy, \PLB{148}{84}{69}.}
\REF\HL{L.L. Chau and W.Y. Keung, \PRL{53}{84}{1802};
        H. Harari and M. Leurer, \PLB{181}{86}{123}.}
\REF\NIR{Y. Nir, SLAC--PUB--5874, September 1992.}
\REF\F{H. Fritzsch, \PLB{73}{78}{317}.}
\REF\SCP{J.E. Kim, \PRT{150}{87}{1}; H.Y. Cheng, \PRT{158}{88}{1}.}

\singlespace
\rl{WIS--93/35/APR--PH}
\rl{\today}
\rl{T}
\normalspace
\smallskip
\titlestyle{\bf{Cabibbo--Kobayashi--Maskawa Mixing in
Superstring Derived Standard--like Models}}
\author{Alon E. Faraggi
{\footnote*{e--mail address: fhalon@weizmann.bitnet. Address after
September 1 1993, School of Natural Sciences, Institute for Advanced Study,
Olden Lane, Princeton NJ, 08540.}}
and Edi Halyo{\footnote\dag{e--mail address: jphalyo@weizmann.bitnet}}}
\smallskip
\centerline {Department of Physics, Weizmann Institute of Science}
\centerline {Rehovot 76100, Israel}
\vskip 4 cm
\titlestyle{\bf ABSTRACT}

We examine the problem of three generation quark
flavor mixing in realistic, superstring derived standard--like models,
constructed in the free fermionic formulation.
We study the sources of family mixing in these models and discuss the
necessary conditions to obtain a realistic Cabibbo--Kobayashi--Maskawa (CKM)
mixing matrix.
In a specific model, we estimate the mixing angles and discuss the weak CP
violating phase.  We argue that the superstring
standard--like models can produce
a realistic CKM mixing matrix. We discuss the possible textures of quark
mass matrices that may be obtained in these models.

\singlespace
\vskip 0.5cm
\nopagenumbers
\pageno=0
\endpage
\normalspace
\pagenumbers

\centerline{\bf 1. Introduction}

The quest of theoretical physics for many years has been to understand the
origin of fermion masses and mixing. The standard model which is consistent
with all experiments to date, uses thirteen free parameters to parametrize
the observed spectrum. Grand Unified Theories (GUTs) and Supersymmetric Grand
Unified Theories (SUSY GUTs) reduce the number of free parameters and are able
to explain inter--family relations between some of the masses.
However, GUTs and SUSY GUTs can explain neither the mass hierarchy among the
generations nor the origin and the amount of observed family mixing.
Within the context of unified theories it is plausible that the number of
generations and the structure of the fermion mass matrices have their origin
in a more fundamental theory at the Planck scale.
Indeed, the best known Planck scale theory, namely superstring theory [\GSW],
indicates that the number of chiral generations is related to the Euler
characteristic of the compactified six dimensional space at the Planck scale.
Therefore, it is important to examine whether realistic superstring
models can lead to a qualitative understanding of the fermion mass matrices.

In Ref. [\EU,\TOP,\SLM] realistic superstring standard--like models
were constructed in the four dimensional free fermionic formulation.
The realistic models in the free fermionic formulation
[\EU,\TOP,\SLM,\REVAMP,\FNY,\ALR] have the attractive
property of correlating the reduction to three generations with the
factorization of the gauge group into observable and hidden sectors, and
with the breaking of unwanted non--Abelian horizontal symmetries in the
observable sector to $U(1)$ factors.
A detailed discussion on the construction of the free fermionic standard--like
models was given in Ref. [\SLM]. The models of interest must possess the
following properties:

\parindent=-15pt

1. The gauge group is $SU(3)_C\times SU(2)_L\times {U(1)^n}\times$hidden,
with $N=1$ space-time supersymmetry.

2. Three generations of chiral fermions
and their superpartners, with the correct quantum numbers
under ${SU(3)_C\times SU(2)_L\times U(1)_Y}$.

3. The spectrum should contain Higgs doublets that can  produce
realistic gauge symmetry breaking.

4. Anomaly cancellation, apart from a single ``anomalous" U(1)
which is  canceled by  application of the
Dine--Seiberg--Witten (DSW) mechanism [\DSW].

\bigskip
\parindent=15pt

The combined constraints (1--4) impose strong restrictions on the
possible boundary condition basis vectors and GSO projection
coefficients, and result in a class of realistic standard--like models
with
unique properties [\SLM].
First, they produce
three and only three generations of chiral fermions. Second, proton decay
from dimension four and five operators is suppressed due to gauged $U(1)$
symmetries and a unique superstringy doublet--triplet splitting mechanism
[\SLM]. Finally, the standard--like models
suggest an explanation for the fermion mass hierarchy [\TOP,\NRT,\FM].
At the tree level of the superpotential only the top quark
gets mass. Mass terms for lighter quarks and leptons are obtained from higher
order nonrenormalizable terms.
The allowed nonrenormalizable terms in the fermion mass matrices are
constrained
by the horizontal symmetries that are derived in the standard--like models.
The horizontal symmetries arise due to the compactification from ten to four
dimensions. In the realistic free fermionic models the horizontal symmetries
reflect the underlying $Z_2\times Z_2$ orbifold compactification [\FM].

An important property of the superstring standard--like models is the
absence of gauge and gravitational anomalies apart from a single ``anomalous
$U(1)$" symmetry. This anomalous $U(1)_A$ generates a Fayet--Iliopoulos term
that breaks supersymmetry at the Planck scale [\DSW]. Supersymmetry is restored
and $U(1)_A$ is broken by giving VEVs to a set of standard model singlets in
the
massless string spectrum along the flat F and D directions [\ADS].
The $SO(10)$ singlet fields in the nonrenormalizable terms
obtain non--vanishing
VEVs by the application of the DSW mechanism. Thus, the order $N$
nonrenormalizable terms, of the form $cffh(\Phi/M)^{N-3}$,
become effective trilinear terms, where $f,h,\Phi$ denote fermions, scalar
doublets and scalar singlets, respectively. $M$ is a Planck scale mass
to be defined later. The effective Yukawa couplings are given by
$\lambda=c(\langle \Phi \rangle/M)^{N-3}$
where the calculable coefficients $c$ are of order one [\KLN].
In this manner quark mass terms, as well as quark mixing terms, can be
obtained. Realistic quark masses and mixing can be obtained for a suitable
choice of scalar VEVs.

In a previous letter [\FH], we studied the mixing between the two lightest
generations.
We showed that for a suitable choice of scalar singlet VEVs, a Cabibbo angle of
the correct order of magnitude can be obtained in standard--like models.
In this paper, we extend our analysis to the case of three generation mixing.
We demonstrate that mixing among three generations and a weak CP violating
phase of the correct order of magnitude can be obtained in the standard--like
models. We illustrate our results in a specific model and
discuss the general properties of our results that are expected to be valid
for a large class of standard--like models.

The paper is organized as follows. In Section 2 we review the
superstring standard--like models.
We discuss
the structure of the massless spectrum and emphasize the general properties
of the standard--like
models that are reflected in the generation mixing. In Section 3,
we obtain the tree level superpotential and the nonrenormalizable terms.
We discuss the form of the generation mixing nonrenormalizable terms
in the standard--like models. We argue that the generation mixing
terms reflect the general structure of the massless spectrum in these models.
In section 4, we discuss the case of two generation Cabibbo mixing.
In section 5, we extend our analysis to the case of three generation
mixing. We discuss the possibility of obtaining mixing angles and weak
CP violating phase of the correct order of magnitude. We present
an F and D flat solution that yields a semi--realistic CKM matrix.
In section 6, we discuss the relation between quark mass matrices in the
standard--like models and ansatze for quark mass matrices.
Our conclusions are summarized in section 7.

\bigskip
\centerline{\bf 2. The superstring standard--like models}

The superstring standard--like models are constructed in the four
dimensional free fermionic formulation [\FFF].
The models are generated by a basis of eight boundary condition vectors
for all world--sheet fermions. The first five vectors in the basis
consist of the NAHE
set $\{{\bf 1},S,b_1,b_2,b_3\}$ [\SLM,\REVAMP].
The first five vectors (including the vector {\bf 1}) in the basis are
$$\eqalignno{S&=({\underbrace{1,\cdots,1}_{{\psi^\mu},
{\chi^{1,...,6}}}},0,\cdots,0
\vert 0,\cdots,0).&(1a)\cr
b_1&=({\underbrace{1,\cdots\cdots\cdots,1}_
{{\psi^\mu},{\chi^{12}},y^{3,...,6},{\bar y}^{3,...,6}}},0,\cdots,0\vert
{\underbrace{1,\cdots,1}_{{\bar\psi}^{1,...,5},
{\bar\eta}^1}},0,\cdots,0).&(1b)\cr
b_2&=({\underbrace{1,\cdots\cdots\cdots\cdots\cdots,1}_
{{\psi^\mu},{\chi^{34}},{y^{1,2}},
{\omega^{5,6}},{{\bar y}^{1,2}}{{\bar\omega}^{5,6}}}}
,0,\cdots,0\vert
{\underbrace{1,\cdots,1}_{{{\bar\psi}^{1,...,5}},{\bar\eta}^2}}
,0,\cdots,0).&(1c)\cr
b_3&=({\underbrace{1,\cdots\cdots\cdots\cdots\cdots,1}_
{{\psi^\mu},{\chi^{56}},{\omega^{1,\cdots,4}},
{{\bar\omega}^{1,\cdots,4}}}},0,\cdots,0
\vert {\underbrace{1,\cdots,1}_{{\bar\psi}^{1,...,5},
{\bar\eta}^3}},0,\cdots,0).&(1d)\cr}$$
with the choice of generalized GSO projections
$$c\left(\matrix{b_i\cr
                                    b_j\cr}\right)=
c\left(\matrix{b_i\cr
                                    S\cr}\right)=
c\left(\matrix{1\cr
                                    1\cr}\right)=-1,\eqno(2)$$
and the others given by modular invariance.

The gauge group after the NAHE set is $SO(10)\times E_8\times SO(6)^3$ with
$N=1$ space--time supersymmetry and 48 spinorial $16$ of $SO(10)$.
The NAHE set is common to all the realistic models in the free fermionic
formulation. The special properties of the NAHE set are
emphasized in Ref. [\SLM].
In short, the vectors $b_1,b_2~ \hbox{and}~b_3$ of the NAHE set perform several
functions. First, they produce the chiral generations. Second, they split the
observable and hidden sectors. Finally, they determine the chirality of the
massless generations. Models based on the NAHE set correspond to models
that are based on
$Z_2\times Z_2$ orbifold compactification with nontrivial background fields.
This correspondence is best illustrated by adding the basis vector
$$X=(0,\cdots,0\vert{\underbrace{1,\cdots,1}_{{{\bar\psi}^{1,\cdots,5}},
{{\bar\eta}^{1,2,3}}}},0,\cdots,0)\eqno(3)$$
to the NAHE set. The gauge group is extended to $E_6\times U(1)^2\times
E_8\times SO(4)^3$ with $N=1$ supersymmetry and 24 chiral 27 of $E_6$.
The same model is obtained in the orbifold language by moding out an $SO(12)$
lattice by a $Z_2\times Z_2$ discrete symmetry with
``standard embedding" [\FM]. The
internal fermionic states $\{y,\omega \vert \bar y,\bar \omega\}$ correspond
to the six left--moving and the six right--moving compactified dimensions in
the orbifold language. In the construction of the standard--like models
beyond the NAHE set, the assignment of boundary conditions to the set of
internal fermions $\{y,\omega \vert \bar y,\bar \omega\}$ determines many of
the properties of the low--energy spectrum such as the number of generations,
the presence of Higgs doublets, Yukawa couplings, etc [\SLM].
We would like to emphasize that many of the low energy properties
are closely related to the $Z_2\times Z_2$ orbifold compactification.
In particular, each of the three chiral generations is obtained from
a distinct twisted sector of the orbifold model. The horizontal symmetries
of each generation correspond to the three orthogonal complex planes
of the $Z_2\times Z_2$ orbifold.

The standard--like models are constructed by adding three additional
vectors to the NAHE set [\EU,\TOP,\SLM,\FNY].
Three additional vectors are needed to reduce the
number of generations to one generation from each sector $b_1$, $b_2$
and $b_3$. The three vectors that extend the NAHE set and the choice
of generalized GSO projection coefficients for our
model are given in Table 1 [\EU].
The observable and hidden gauge groups after application
of the generalized GSO projections are
$SU(3)_C\times U(1)_C\times
 SU(2)_L\times U(1)_L\times U(1)^6${\footnote*{
$U(1)_C={3\over 2}U(1)_{B-L}$ and
$U(1)_L=2U(1)_{T_{3_R}}$.}}
and $SU(5)_H\times SU(3)_H\times U(1)^2$, respectively.
The weak hypercharge is given by
$U(1)_Y={1\over 3}U(1)_C + {1\over 2}U(1)_L$ and has the standard $SO(10)$
embedding. The orthogonal
combination is given by $U(1)_{Z^\prime}= U(1)_C - U(1)_L$.
The vectors $\alpha,\beta,\gamma$ break the $SO(6)_{j}$
horizontal symmetries to
$U(1)_{r_j}\times U(1)_{r_{j+3}}$ ($j=1,2,3$),
which correspond to the right--moving world--sheet
currents ${\bar\eta}^j_{1\over2}{{\bar\eta}^{j^*}}_{1\over2}$
($j=1,2,3$) and
${{\bar y}^3{\bar y}^6,{\bar y}^1{\bar\omega}^5,
{\bar\omega}^2{\bar\omega}^4}$, respectively.
For every right--moving $U(1)_r$ gauge symmetry there is
a left--moving global $U(1)_\ell$ symmetry. The first three
correspond to the charges of the supersymmetry generator
$\chi^{12}$, $\chi^{34}$ and $\chi^{56}$. The last three,
$U(1)_{\ell_{j+3}}$ $(j=1,2,3)$,
correspond to the complexified left--moving fermions
$y^3y^6$, $y^1\omega^5$ and $\omega^2\omega^4$.
Finally, the model contains six Ising model operators
that are obtained by pairing a left--moving
real fermion with a right--moving real fermion,
$\sigma^i_\pm=\{\omega^1{\bar\omega}^1,
y^2{\bar y}^2, \omega^3{\bar\omega}^3, y^4{\bar y}^4,
y^5{\bar y}^5, \omega^6{\bar\omega}^6\}_\pm$.

The basis vectors span a finite additive group $\Xi=Z_2^7 \times Z_4$.
A general property of the free fermionic models, which are based on the
NAHE set and that
use a $Z_4$ twist to break the gauge symmetry from $SO(2n)$ to
$SU(n) \times U(1)$,
is the presence of the sectors $b_j$ and $b_j+2\gamma+(I)$
$j=(1,2,3)$ in the massless spectrum. The sectors $b_j$ produce the chiral
generations and the sectors $b_j+2\gamma+(I)$ produce representations of the
hidden gauge groups that are $SO(10)$ singlets with horizontal charges.
The vector $2\gamma$, in effect, when added to the NAHE set, plays the role
of the vector $X$ in Eq. (3). It splits the $\{{\bar y},{\bar\omega}\}$
right--moving fermionic states from ${\bar\eta}_{_{1,2,3}}$, and breaks the
horizontal symmetries from $SO(6)_j$ to $SO(4)_j\times U(1)_j$. However,
rather than enhancing the observable gauge group from $SO(10)$ to $E_6$,
it breaks the hidden gauge group from $E_8$ to $SO(16)$ and produces
massless states in the vector representation of $SO(16)$, from the sectors
$b_j+2\gamma$.
We will argue that this structure of the additive group, in these models, is
the essential feature behind the generation mixing.

The full massless spectrum was presented in Ref. [\EU].
Here we list only the states that are relevant for the quark mass matrices.
The following massless states are produced by the sectors $b_{1,2,3}$,
$S+b_1+b_2+\alpha+\beta$, $O$ and their superpartners in the observable
sector:

(a) The $b_{1,2,3}$ sectors produce three $SO(10)$ chiral generations,
$G_\alpha=e_{L_\alpha}^c+u_{L_\alpha}^c+N_{L_\alpha}^c+d_{L_\alpha}^c+
Q_\alpha+L_\alpha$ $(\alpha=1,\cdots,3)$ where
$$\eqalignno{{e_L^c}&\equiv [(1,{3\over2});(1,1)];{\hskip .6cm}
{u_L^c}\equiv [({\bar 3},-{1\over2});(1,-1)];{\hskip .2cm}
Q\equiv [(3,{1\over2});(2,0)]{\hskip 2cm}
&(4a,b,c)\cr
{N_L^c}&\equiv [(1,{3\over2});(1,-1)];{\hskip .2cm}
{d_L^c}\equiv [({\bar 3},-{1\over2});(1,1)];{\hskip .6cm}
L\equiv [(1,-{3\over2});(2,0)]{\hskip 2cm}
&(4d,e,f)\cr}$$
of $SU(3)_C\times U(1)_C\times SU(2)_L\times U(1)_L$, with charges under the
six horizontal $U(1)$s. We obtain from the sector $b_1$
$$({e_L^c}+{u_L^c})_{{1\over2},0,0,{1\over2},0,0}+
({d_L^c}+{N_L^c})_{{1\over2},0,0,{-{1\over2}},0,0}+
(L)_{{1\over2},0,0,{1\over2},0,0}+(Q)_{{1\over2},0,0,-{1\over2},0,0},
\eqno(5a)$$
the sector $b_2$,
$$({e_L^c}+{u_L^c})_{0,{1\over2},0,0,{1\over2},0}+
({N_L^c}+{d_L^c})_{0,{1\over2},0,0,-{1\over2},0}+
(L)_{0,{1\over2},0,0,{1\over2},0}+
(Q)_{0,{1\over2},0,0,-{1\over2},0},
\eqno(5b)$$
and the sector $b_3$,
$$({e_L^c}+{u_L^c})_{0,0,{1\over2},0,0,{1\over2}}+
({N_L^c}+{d_L^c})_{0,0,{1\over2},0,0,-{1\over2}}+
(L)_{0,0,{1\over2},0,0,{1\over2}}+(Q)_{0,0,{1\over2},0,0,-{1\over2}}.
\eqno(5c)$$
The vectors $b_1,b_2,b_3$ are the only vectors in the additive group
$\Xi$ which give rise to spinorial $16$ of $SO(10)$.

(b) The ${S+b_1+b_2+\alpha+\beta}$ sector gives
$$\eqalignno{h_{45}&\equiv{[(1,0);(2,1)]}_
{-{1\over2},-{1\over2},0,0,0,0} {\hskip .5cm}
D_{45}\equiv{[(3,-1);(1,0)]}_
{-{1\over2},-{1\over2},0,0,0,0}&(6a,b)\cr
\Phi_{45}&\equiv{[(1,0);(1,0)]}_
{-{1\over2},-{1\over2},-1,0,0,0}  {\hskip .5cm}
\Phi^{\pm}_1\equiv{[(1,0);(1,0)]}_
{-{1\over2},{1\over2},0,\pm1,0,0}&(6c,d)\cr
\Phi^{\pm}_2&\equiv{[(1,0);(1,0)]}_
{-{1\over2},{1\over2},0,0,\pm1,0} {\hskip .5cm}
\Phi^{\pm}_3\equiv{[(1,0);(1,0)]}_
{-{1\over2},{1\over2},0,0,0,\pm1}&(6e,f)\cr}$$
(and their conjugates ${\bar h}_{45}$, etc.).
The states are obtained by acting on the vacuum
with the fermionic oscillators
${\bar\psi}^{4,5},{\bar\psi}^{1,...,3},{\bar\eta}^3,{\bar y}^3\pm
i{\bar y}^6,{\bar y}^1\pm{i{\bar\omega}^5},
{\bar\omega}^2{\pm}i{\bar\omega}^4$,
respectively  (and their complex conjugates for ${\bar h}_{45}$, etc.).

(c) The Neveu--Schwarz $O$ sector gives, in addition to  the graviton,
dilaton, antisymmetric tensor and spin 1 gauge bosons,  scalar
electroweak doublets and singlets:
$$\eqalignno{{h_1}&\equiv{[(1,0);(2,-1)]}_{1,0,0,0,0,0}
{\hskip 2cm}\Phi_{23}\equiv{[(1,0);(1,0)]}_{0,1,-1,0,0,0}&(7a,b)\cr
{h_2}&\equiv{[(1,0);(2,-1)]}_{0,1,0,0,0,0}
{\hskip 2cm}\Phi_{13}\equiv{[(1,0);(1,0)]}_{1,0,-1,0,0,0}&(7c,d)\cr
{h_3}&\equiv{[(1,0);(2,-1)]}_{0,0,1,0,0,0}
{\hskip 2cm}\Phi_{12}\equiv{[(1,0);(1,0)]}_{1,-1,0,0,0,0}&(7e,f)\cr}$$
(and their conjugates ${\bar h}_1$, etc.).
Finally, the Neveu--Schwarz sector gives rise to three singlet
states that are neutral under all the U(1) symmetries.
$\xi_{1,2,3}:{\hskip .2cm}{\chi^{12}_{1\over2}{\bar\omega}^3_{1\over2}
{\bar\omega}^6_{1\over2}{\vert 0\rangle}_0},$
 ${\chi^{34}_{{1\over2}}{\bar y}_{1\over2}^5{\bar\omega}_{1\over2}^1
{\vert 0\rangle}_0},$
 $\chi^{56}_{1\over2}{\bar y}_{1\over2}^2{\bar y}_{1\over2}^4
{\vert 0\rangle}_0.$

The sectors $b_i+2\gamma+(I){\hskip .2cm} (i=1,..,3)$ give vector--like
representations that are
$SU(3)_C\times SU(2)_L\times {U(1)_L}\times {U(1)_C}$
singlets and transform as $5$, ${\bar 5}$ and $3$, ${\bar 3}$
under the
hidden $SU(5)$ and $SU(3)$ gauge groups, respectively (see Table 2).
As will be shown below, the states from the sectors
$b_j+2\gamma$ produce the mixing between the chiral generations. We
would like to emphasize that the structure of the massless spectrum
exhibited in Eqs. (5--7), and in Table 2, is common to a large number
of free fermionic standard--like models. All the standard--like models
contain three chiral generations from the sectors $b_j$, vector--like
representations from the sectors $b_j+2\gamma$, and Higgs doublets from
the Neveu--Schwarz sector.
The vector combination of
$\alpha+\beta$ plus some combination of $\{b_1,b_2,b_3\}$, produces
additional doublets and singlets, and exists in the models that were
found to admit F and D flat solution [\EU,\TOP,\SLM], but not in the
model Ref. [\FNY]. We will show that mixing terms are obtained
in all these models.
We will argue that
the source of the family mixing is a general characteristic of these
models. It arises due to the basic set $\{{\bf 1},S,b_1,b_2,b_3\}$
and the use of the $Z_4$ twist to break the symmetry
from $SO(2n)$ to $SU(n)\times U(1)$.

In addition to the states above, the massless spectrum contains massless
states from sectors with some combination of
$\{b_1,b_2,b_3,\alpha,\beta\}$ and $\gamma+(I)$. These states are
model dependent and carry either fractional electric charge or
$U(1)_{Z^\prime}$ charge. As argued in Ref. [\NRT,\FHn] the $U(1)_{Z^\prime}$
symmetry has to be broken at an intermediate energy scale that is
suppressed relative to the Planck scale. Therefore, the states from these
sectors do not play a significant role in the quark mass matrices and
we do not consider them in this paper.

The model contains six anomalous $U(1)$ gauge symmetries:
Tr${U_1}=24$, Tr${U_2}=24$, Tr${U_3}=24$,
Tr${U_4}=-12$, Tr${U_5}=-12$, Tr${U_6}=-12$.
Of the six anomalous $U(1)$s,  five can be rotated by
an orthogonal transformation and one combination remains anomalous.
The six orthogonal combinations are given by [\EU],
$$\eqalignno{{U^\prime}_1&=U_1-U_2{\hskip .5cm},{\hskip .5cm}
{U^\prime}_2=U_1+U_2-2U_3,&(8a,b)\cr
{U^\prime}_3&=U_4-U_5{\hskip .5cm},{\hskip .5cm}
{U^\prime}_4=U_4+U_5-2U_6,&(8c,d)\cr
{U^\prime}_5&=U_1+U_2+U_3+2U_4+2U_5+2U_6,&(8e)\cr
U_A&=2U_1+2U_2+2U_3-U_4-U_5-U_6,&(8f)\cr}$$
with $Tr(Q_A)=180.$
The set of F and D constraints is given by the following equations:
$$\eqalignno{&D_A=\sum_k Q^A_k \vert \chi_k \vert^2={-g^2e^{\phi_D}
\over 192\pi^2}Tr(Q_A) &(9a) \cr
&D^{\prime j}=\sum_k Q^{\prime j}_k \vert \chi_k \vert^2=0 \qquad j=1
\ldots 5 &(9b) \cr
&D^j=\sum_k Q^j_k \vert \chi_k \vert^2=0 \qquad j=C,L,7,8 &(9c) \cr
&W={\partial W\over \partial \eta_i} =0 &(9d) \cr}$$
where $\chi_k$
are the fields that
get VEVs
and $Q^j_k$ are their charges. $W$ is the tree level
superpotential.

\bigskip
\centerline{\bf 3. The superpotential and mixing terms}

We now turn to the superpotential of the model. Trilinear and nonrenormalizable
contributions to the superpotential are obtained by calculating correlators
between vertex operators [\KLN]
$$A_N\sim\langle V_1^fV_2^fV_3^b\cdot\cdot\cdot V_N^b\rangle \eqno(10)$$
where $V_i^f$ $(V_i^b)$ are the fermionic (scalar)
components of the vertex operators. The non--vanishing terms are obtained by
applying the rules of Ref. [\KLN]. In order to obtain the correct ghost charge,
some of the vertex operators are picture changed by taking
$$V_{q+1}(z)=\lim_{w \to z}e^c(w)T_F(w)V_q(z) \eqno(11)$$
where $T_F$ is the world--sheet super current given by
$$T_F=\psi^{\mu}\partial_{\mu}X+i\sum_{I=1}^6 \chi^I y^I \omega^I=T_F^0+
T_F^{-1}+T_F^{+1} \eqno(12)$$
with
$$T_F^{-1}=e^{-i\chi^{12}}\tau_{_{12}}+e^{-i\chi^{34}}\tau_{_{34}}+
e^{-i\chi^{56}}\tau_{_{56}}  \qquad  T_F^{-1}=(T_F^{+1})^* \eqno(13)$$
where $\tau_{ij}={i\over \sqrt 2}(y^i\omega^i+iy^j\omega^j)$ and
$e^{i\chi^{ij}}={1\over \sqrt2}(\chi^i+i\chi^j)$.

Several observations simplify the analysis of the potential
non--vanishing terms.
First, it is seen that only the $T_F^{+1}$ piece of $T_F$
contributes to $A_N$ [\KLN].
Second, in the standard--like models [\EU,\TOP] the pairing of left--moving
fermions is $y^1\omega^5,\omega^2\omega^4~ \hbox{and}~ y^3y^6$. One of the
fermionic states in every term $y^i\omega^i$ ($i=1,\ldots,6$) is complexified
and therefore can be written, for example for $y^3$ and $y^6$, as
$$y^3={1\over \sqrt2}(e^{iy^3y^6}+e^{-iy^3y^6}),\quad y^6={1\over \sqrt2}
(e^{iy^3y^6}-e^{-iy^3y^6}). \eqno(14)$$
Consequently, every picture changing operation changes the total $U(1)_\ell=
U(1)_{\ell_4}+U(1)_{\ell_5}+U(1)_{\ell_6}$
charge by $\pm 1$. An odd (even) order term
requires an even (odd) number of picture changing operations to get the
correct ghost number [\KLN]. Thus, for $A_N$ to be non--vanishing, the total
$U(1)_\ell$
charge, before picture changing, has to be an even (odd) number for even (odd)
order terms. Similarly, in every pair $y^i\omega^i$, one real fermion, either
$y^i$ or $\omega^i$, remains real and is paired with the corresponding
right--moving real fermion to form an Ising model operator. Every picture
changing operation changes the number of left--moving real fermions by one.
This property of the standard--like models  [\EU,\TOP] significantly
reduces the number of potential non--vanishing terms.

At the cubic level the following terms are obtained in the observable
sector [\EU],
$$\eqalignno{W_3&=\{(
{u_{L_1}^c}Q_1{\bar h}_1+{N_{L_1}^c}L_1{\bar h}_1+
{u_{L_2}^c}Q_2{\bar h}_2+{N_{L_2}^c}L_2{\bar h}_2+
{u_{L_3}^c}Q_3{\bar h}_3+{N_{L_3}^c}L_3{\bar h}_3)\cr
&\qquad
+{{h_1}{\bar h}_2{\bar\Phi}_{12}}
+{h_1}{\bar h}_3{\bar\Phi}_{13}
+{h_2}{\bar h}_3{\bar\Phi}_{23}
+{\bar h}_1{h_2}{\Phi_{12}}
+{\bar h}_1{h_3}{\Phi_{13}}
+{\bar h}_2{h_3}{\Phi_{23}}
+\Phi_{23}{\bar\Phi}_{13}{\Phi}_{12}\cr
&\qquad
+{\bar\Phi}_{23}{\Phi}_{13}{\bar\Phi}_{12}
+{\bar\Phi}_{12}({\bar\Phi}_1^+{\bar\Phi}_1^-
+{\bar\Phi}_2^+{\bar\Phi}_2^-
+{\bar\Phi}_3^+{\bar\Phi}_3^-)
+{\Phi_{12}}(\Phi_1^-\Phi_1^+
+\Phi_2^-\Phi_2^+
+\Phi_3^-\Phi_3^+)\cr
&\qquad
+{1\over2}\xi_3(\Phi_{45}{\bar\Phi}_{45}
+h_{45}{\bar h}_{45}
+D_{45}{\bar D}_{45}+\Phi_1^+{\bar\Phi}_1^++
\Phi_1^-{\bar\Phi}_1^-+\Phi_2^+{\bar\Phi}_2^++\Phi_2^-{\bar\Phi}_2^-
+\Phi_3^+{\bar\Phi}_3^+
\cr
&\qquad
+\Phi_3^-{\bar\Phi}_3^-)
+h_3{\bar h}_{45}\Phi_{45}+{\bar h}_3h_{45}{\bar\Phi}_{45}\}\quad&(15)\cr}$$
with a common normalization constant ${\sqrt 2}g$.

It is seen that only Yukawa couplings of the $+{2\over3}$ charged
quarks and neutral leptons
appear in the tree level superpotential. This is a result of our choice of
the basis vector $\gamma$ given in Table 2 [\YUKAWA].
In the analysis of nonrenormalizable terms we impose the F--flatness
restriction $\langle{\bar\Phi}_{12},\Phi_{12},\xi_3\rangle\equiv0$
[\NRT]. In addition, we take $\langle \Phi_{23},\bar \Phi_{45} \rangle=0$.
At the cubic level there are two pairs of light Higgs doublets, which
are combinations of $\{h_1,h_2,h_{45}\}$ and
$\{{\bar h}_1,{\bar h}_2,{\bar h}_{45}\}$ [\NRT,\FM].
One additional pair receives heavy mass at the intermediate scale of
$U(1)_{Z^\prime}$ breaking. The light Higgs representations, below this
scale, may consist of $h_{45}$ and a combination of ${\bar h}_1$ or
${\bar h}_2$ and ${\bar h}_{45}$.
As the mixing is dominantly in the down quark
sector, we do not loose any generality by assuming
the light Higgs representation
to be ${\bar h}_1$ and $h_{45}$.
Since the heavy Higgs
doublets decouple at low energies, only the Yukawa couplings with $\bar h_i$
remain in the superpotential given by Eq. (15). Therefore only the top quark
gets
a tree level mass.
The other quarks get their masses from
higher order nonrenormalizable terms which contain the light Higgs doublets.

At the quartic order there are no potential quark mass terms. At the
quintic order the following mass terms are obtained,
$$\eqalignno{&d_1Q_1h_{45}{\Phi}_1^+\xi_2
  {\hskip 2cm}d_2Q_2h_{45}{\bar\Phi}_2^-\xi_1 &(16a,b)\cr
             &u_1Q_1({\bar h}_{45}\Phi_{45}{\bar\Phi}_{13}+
  {\bar h}_2{\Phi}_i^+{\Phi}_i^-)&(16c)\cr
      &u_2Q_2({\bar h}_{45}\Phi_{45}{\bar\Phi}_{23}+
  {\bar h}_1{\bar\Phi}_i^+{\bar\Phi}_i^-)&(16d)\cr
      &(u_1Q_1h_1+u_2Q_2h_2)
  {{\partial W}\over{\partial\xi}_3}.&(16e)\cr}$$
At order $N=6$ we obtain mixing terms for $-{1\over3}$ charged quarks,
$$\eqalignno{
&d_3Q_2h_{45}\Phi_{45}V_3{\bar V_2},{\hskip 2cm}
d_2Q_3h_{45}\Phi_{45}V_2{\bar V_3},&(17a,b)\cr
&d_3Q_1h_{45}\Phi_{45}V_3{\bar V_1},{\hskip 2cm}
d_1Q_3h_{45}\Phi_{45}V_1{\bar V_3},&(17c,d)\cr}$$
At order $N=7$ we obtain in the down quark sector,
$$\eqalignno{
&d_2Q_1h_{45}\Phi_{45}(V_1{\bar V_2}+V_2{\bar V_1})\xi_i{\hskip .8cm}
d_1Q_2h_{45}\Phi_{45}(V_1{\bar V_2}+V_2{\bar V_1})\xi_i&(18a,b)\cr
&d_1Q_3h_{45}\Phi_{45}V_3{\bar V_1}\xi_2{\hskip 2.5cm}
d_3Q_1h_{45}\Phi_{45}V_1{\bar V_3}\xi_2&(18c,d)\cr
&d_2Q_3h_{45}\Phi_{45}V_3{\bar V_2}\xi_1{\hskip 2.5cm}
d_3Q_2h_{45}\Phi_{45}V_2{\bar V_3}\xi_1,&(18e,f) \cr}$$
where $\xi_i=\{\xi_1,\xi_2\}$.
In the up quark sector we obtain,
$$\eqalignno{
&u_1Q_2{\bar h}_1\Phi_{45}\{{\bar\Phi}_2^-(T_1{\bar T_2}+T_2{\bar T_1})+
                          {\bar\Phi}_1^+(V_1{\bar V_2}+V_2{\bar V_1}\}&(19a)\cr
&u_2Q_1{\bar h}_1\Phi_{45}\{{\bar\Phi}_1^-(T_1{\bar T_2}+T_2{\bar T_1})+
                          {\bar\Phi}_2^+(V_1{\bar V_2}+V_2{\bar V_1}\}&(19b)\cr
&u_1Q_2{\bar h}_2\Phi_{45}\{{\Phi}_2^+(T_1{\bar T_2}+T_2{\bar T_1})+
                            {\Phi}_1^-(V_1{\bar V_2}+V_2{\bar V_1}\}&(19c)\cr
&u_2Q_1{\bar h}_2\Phi_{45}\{{\Phi}_1^+(T_1{\bar T_2}+T_2{\bar T_1})+
                            {\Phi}_2^-(V_1{\bar V_2}+V_2{\bar V_1}\}&(19d)\cr
&u_3Q_1{\bar h}_1\Phi_{45}\{{\bar\Phi}_1^-T_1{\bar T_3}+
                            {\bar\Phi}_3^+V_3{\bar V_1}\}{\hskip .5cm}
 u_1Q_3{\bar h}_1\Phi_{45}\{{\bar\Phi}_3^-T_1{\bar T_3}+
                            {\bar\Phi}_1^+V_3{\bar V_1}\}&(19e)\cr
&u_3Q_1{\bar h}_2\Phi_{45}\{{\Phi}_3^+T_1{\bar T_3}+
                            {\Phi}_1^-V_3{\bar V_1}\}{\hskip .5cm}
 u_1Q_3{\bar h}_2\Phi_{45}\{{\Phi}_3^+T_1{\bar T_3}+
                            {\Phi}_1^-V_3{\bar V_1}\}&(19f)\cr
&u_3Q_2{\bar h}_1\Phi_{45}\{{\bar\Phi}_2^-T_2{\bar T_3}+
                            {\bar\Phi}_3^+V_3{\bar V_2}\}{\hskip .5cm}
 u_2Q_3{\bar h}_1\Phi_{45}\{{\bar\Phi}_3^-T_2{\bar T_3}+
                            {\bar\Phi}_2^+V_3{\bar V_2}\}&(19g)\cr
&u_3Q_2{\bar h}_2\Phi_{45}\{{\Phi}_2^+T_2{\bar T_3}+
                            {\Phi}_3^-V_3{\bar V_2}\}{\hskip .5cm}
 u_2Q_3{\bar h}_2\Phi_{45}\{{\Phi}_3^-T_2{\bar T_3}+
                            {\Phi}_2^-V_3{\bar V_2}\}&(19h)\cr}$$
At order $N=7$ we obtain generation mixing terms in the up and down quark
sectors. The states that induce the mixing come from the
sectors $b_j+2\gamma$. In the up quark sector, mixing is obtained
by $5$, ${\bar 5}$ and $3$, ${\bar 3}$ of the hidden $SU(5)$ and $SU(3)$
gauge groups, respectively. In the down quark sector, the mixing is only by
the $3$, ${\bar 3}$ of the hidden $SU(3)$ gauge groups.
At order $N=8$ we obtain mixing in the
down quark sector
by the $SU(5)$ states from the sectors $b_j+2\gamma$,
$$\eqalignno{
&d_3Q_1h_{45}\Phi_{45}\{{\Phi_1^+}{\bar\Phi}_3^-+
             {\Phi_3^+}{\bar\Phi}_1^-\}T_1{\bar T_3}{\hskip .5cm}
d_1Q_3h_{45}\Phi_{45}\{{\Phi_1^+}{\bar\Phi}_3^-+
             {\Phi_3^+}{\bar\Phi}_1^-\}T_3{\bar T_1}{\hskip 2mm}&(20a)\cr
&d_3Q_2h_{45}\Phi_{45}\{{\Phi_2^+}{\bar\Phi}_3^-+
             {\Phi_3^+}{\bar\Phi}_2^-\}T_2{\bar T_3}{\hskip .5cm}
d_2Q_3h_{45}\Phi_{45}\{{\Phi_2^+}{\bar\Phi}_3^-+
             {\Phi_3^+}{\bar\Phi}_2^-\}T_3{\bar T_2}{\hskip 2mm}&(20b) \cr}$$
The analysis of the nonrenormalizable terms up to order $N=8$
shows that family mixing terms are obtained for all generations.
Before the spontaneous symmetry breaking due to the scalar VEVs, there is no
mixing because of the six generational gauge $U(1)_r$ and the six global
$U(1)_\ell$
symmetries. In general the set of scalar VEVs break all of these symmetries
and induce generation mixing by higher order nonrenormalizable terms.

We observe that the mixing arises due to the states from the sectors
$b_j+2\gamma$.
These sectors, and their relation to the sectors $b_j$, is a general
characteristic of the realistic free fermionic models that use a $Z_4$ twist.
The mixing terms are of the form $f_if_jh\phi^n$, where $f_i,f_j$ are
fermion states from the sectors $b_i,b_j$ with $i\ne j$, $h$ are the light
Higgs representations and $\phi^n$ is a string of standard model singlets.
The fermion states from each sector $b_j$ carry
$U(1)_{\ell_{j+3}}=\pm{1\over 2}$.
The singlets from the NS sector and the sector $b_1+b_2+\alpha+\beta$ all have
$U(1)_{\ell_{j+3}}=0$. Every picture changing operation changes the total
$U(1)_\ell=U(1)_{\ell_4}+U(1)_{\ell_5}+U(1)_{\ell_6}$
by $\pm1$. Thus, in order to
construct nonrenormalizable terms which are invariant under $U(1)_\ell$,
we must tag to $f_if_jh$ additional fields with
$U(1)_{\ell_{j+3}}=\pm{1\over 2}$.
We observe that the only
available states are from the sectors $b_j+2\gamma$,
which are in the fundamental
representations of the hidden gauge group.
Therefore, the family mixing due to these states is a general
characteristic of these models.

We now comment on quark flavor mixing in other standard--like models. The
terms in Eqs. (16--20) were obtained in the model of Ref. [\EU] (model 1)
and similar terms are obtained in the model of Ref. [\TOP] (model 2).
The symmetries of these two models are the same and both contain the
sector $b_1+b_2+\alpha+\beta$ in the massless spectrum. In the observable
sector,
models 1 and 2 differ by the $U(1)_{{\{\ell,r\}}_{j+3}}$ charges of the
massless states from the sector $b_j$.
Consequently, in model 1 [\EU] bottom quark and tau lepton Yukawa couplings
are obtained at the quintic order, while in model 2 [\TOP] they
are obtained at the quartic order [\SLM]. The nonrenormalizable terms,
Eqs. (16--20) may suggest that the condition of realistic quark flavor mixing
requires the presence of the sector $b_1+b_2+\alpha+\beta$ in the
massless spectrum (or the presence of the scalar $\Phi_{45}$). However, by
examining the quark flavor mixing terms
in the model of Ref. [\FNY] (model 3), we obtain at order $N=6$ the
non--vanishing terms ( in the notation of
Ref. [\FNY])
$$\eqalignno{
&d_2Q_1(h_{2}\Phi_{13}+h_3{\bar\Phi}_{12})V_2V_{11},{\hskip 1cm}
 d_1Q_2(h_{2}\Phi_{13}+h_3{\bar\Phi}_{12})V_1V_{12},&(21a,b)\cr
&d_3Q_1(h_{2}\Phi_{13}+h_3{\bar\Phi}_{12})V_2V_{21},{\hskip 1cm}
 d_1Q_3(h_{2}\Phi_{13}+h_3{\bar\Phi}_{12})V_1V_{22},&(21c,d)\cr
&d_3Q_2(h_{2}\Phi_{13}+h_3{\bar\Phi}_{12})\{V_{12}V_{21}+V_{13}V_{24}\},
&(21e)\cr
&d_2Q_3(h_{2}\Phi_{13}+h_3{\bar\Phi}_{12})\{V_{11}V_{22}+V_{14}V_{23}\}.
&(21f)\cr}$$
The states $V_i$ are the states from the sectors $b_j+2\gamma$ in the model
of Ref. [\FNY]. The terms in Eqs. (21) reflect
the dependence of the mixing terms on the interplay between the sectors
$b_j$ and the sectors $b_j+2\gamma$, without the presence of a sector of
the form $\alpha+\beta$ in the massless spectrum.

\bigskip
\centerline{\bf 4. Cabibbo mixing}

In a previous letter [\FH], we showed that there are solutions to the F and D
constraints which give non--negligible Cabibbo mixing between two generations.
In principle generation mixing can arise from two different sources. The first
one is due to condensates of of the states which are in the vector
representation of the hidden gauge group (see Table 2). $T_i$ and $V_i$
which transform as
5's and 3's under $SU(5)_H$ and $SU(3)_H$ form condensates when these gauge
groups get strong, i.e. when
$$\alpha_H(\Lambda_H)={\alpha_H(M)\over{1-(b/{2\pi})\alpha_H(M)
\ln(\Lambda_H/M)}} \sim 1\eqno(22)$$
where $b=(n_f/2)-3N$ and $\alpha_H(M) \sim 0.06$ [\GCU]. The value of
the scalar condensates $\langle \bar V_i V_i \rangle$ or $\langle \bar T_i T_i
\rangle$ is $\sim \Lambda_H^2$, where
$\Lambda_H$ is given by Eq. (22). In our model, for the matter content of
the hidden gauge
groups, $\Lambda_H$ turns out to be too small to give appreciable Cabibbo
mixing. Even for the largest possible hidden gauge group $SU(7)_H$, $\Lambda_7$
turns out to be an order of magnitude too small.

An alternative way to obtain mixing is by giving VEVs to vector representations
of the hidden gauge groups by the F and D constraints given by Eq. (9).
These VEVs will
necessarily break the hidden gauge groups spontaneously. By choosing an
appropriate solution one can easily get
non--negligible mixing. In Ref. [\FH], we
considered a solution to the F and D constraints with the following set of
non--vanishing VEVs:
$$\{V_2,{\bar V}_3,\Phi_{45},\Phi_{23},{\bar\Phi}_{23},\Phi_{13},
{\bar\Phi}_{13},\Phi_1^+,\Phi_2^\pm,{\bar\Phi}_1^\pm,
{\bar\Phi}_2^\pm\,\xi_1,\xi_2\},\eqno(23)$$
with
$$\eqalignno{&\vert\l{V_2}\r\vert^2=\vert\l{\bar V}_3\r\vert^2={1\over5}
\vert\l\Phi_{45}\r\vert^2=\vert\l{\bar\Phi}_1^-\r\vert^2={{g^2\over{16\pi^2}}
{1\over 2\alpha^\prime}} &(24a)\cr
&3\vert\l{\Phi_2^+}\r\vert^2=3\vert\l{{\bar\Phi}_2^+}\r\vert^2=
\vert\l{\Phi_2^-}\r\vert^2=\vert\l{{\bar\Phi}_2^-}\r\vert^2=
\vert\l{\bar\Phi}_{13}\r\vert^2 &(24b)\cr
&{1\over4}\vert\l{\Phi_1^+}\r\vert^2={1\over4}\vert\l{{\bar\Phi}_1^+}\r\vert^2=
\vert\l{\bar\Phi}_{13}\r\vert^2 &(24c)\cr
&\vert\l{\Phi_{23}}\r\vert^2=\vert\l{\bar\Phi}_{23}\r\vert^2=
{1\over3}\vert\l{\bar\Phi}_{13}\r\vert^2 &(24d)\cr
&\vert\l\Phi_{13}\r\vert^2=\vert\l{\bar\Phi}_{13}
\r\vert^2-{{g^2\over 8\pi^2}}{1\over {2\alpha^\prime}} &(24e) \cr}$$
The VEVs of $\xi_1$, $\xi_2$ and ${\bar\Phi}_{13}$
are undetermined and remain free parameters to be fixed.
For this solution, the up mass matrix $M_U$ is diagonal
$$M_u=\hbox{diag}(0,\l\bar \Phi_i^+ \bar \Phi_i^-\r/M^2,1)v_1  \eqno(25)$$
where $v_1=\l \bar h_1 \r$ and the down mass matrix $M_D$ is given by
$$M_d\sim\left(\matrix
{&0
&{{V_2{\bar V}_3\Phi_{45}}\over{M^3}} &0\cr
&{{V_2{\bar V}_3\Phi_{45}\xi_1}\over{M^4}}
&{{{\bar\Phi}_2^-\xi_1}\over{M^2}} &0 \cr
&0 &0
&{{\Phi_1^+\xi_2}\over{M^2}}\cr}\right)v_2\eqno(26)$$
where $v_2=\l h_{45} \r$ and we have used
${1\over2}g\sqrt{2\alpha^\prime}=\sqrt{8\pi}/M_{Pl}$, to define
$M\equiv M_{Pl}/2\sqrt{8\pi}\approx 1.2\times 10^{18}GeV$ [\KLN].
We use the undetermined VEVs of $\bar \Phi_{13}$ and $\xi_2$ to fix $m_b$ and
$m_s$ such that $\l \xi_1 \r \sim M$. We also take $tan \beta=v_1/v_2
\sim 1.5$.
Substituting the values of the VEVs above and
diagonalizing $M_D$ by a biunitary transformation we obtain the
Cabibbo mixing matrix
$$\vert V \vert\sim \left(\matrix {0.98&0.2&0 \cr
                                        0.2&0.98&0 \cr
                                        0&0&1 \cr } \right) \eqno(27)$$
Since the running from the scale $M$ down to the weak scale does not affect
the Cabibbo angle by much [\GLT], we conclude that realistic mixing of the
correct order of magnitude can be obtained in this scenario.

A Cabibbo angle of the correct order of magnitude is obtained due to the
non--vanishing VEVs of $V_2$ and ${\bar V}_3$ along the F and D flat
directions,
Eq. (24). From the nonrenormalizable terms, Eqs. (16--20),
and the F and D flat solution, Eqs. (23,24), we conclude that mixing
between the other generations can be obtained by giving a non--vanishing VEV
to a state from the sector $b_1+2\gamma$. Thus, in order to obtain realistic
CKM mixing matrices, we must find F and D flat solutions with a non--vanishing
VEV for at least one state from each sector $b_j+2\gamma$
(j=1,2,3).

\bigskip
\centerline{\bf 5. KM mixing among three generations}

In this section we consider the mixing between three generations obtained
from the VEVs of hidden sector states. This can be accomplished
by giving VEVs to one hidden sector state from each sector $b_i+2\gamma$.
We impose several conditions
on the VEVs that solve the F and D constraints.
The VEVs should generate mass terms of the correct order of magnitude
for the charm, strange, bottom and top quarks.
This means
that $\Phi_1^+$, $\bar \Phi_2^-$, $\Phi_i^+$, $\Phi_i^-$, $\xi_1$ and $\xi_2$
should get VEVs of the required magnitude. The light Higgs representations
should include $h_{45}$ and one $\bar h_i$. This imposes vanishing VEVs on
$\Phi_{12}$, $\bar \Phi_{12}$, $\bar \Phi_{45}$, $\Phi_{23}$ and $\xi_3$.
We allow a non--vanishing VEV only for one $V_i$ or ${\bar V}_i$ from
every sector $b_i+2\gamma$. This guarantees that terms of the form
$h{\bar h}V_i{\bar V}_i\l\phi\r^n$ will not render all the Higgs doublets
superheavy and cause problems with electroweak--weak symmetry breaking. For
illustrative purposes, we require
mixing terms in the down and up quark mass matrices. Therefore, in
addition to the above fields also $\Phi_{45}$, $\bar \Phi^+_{1,2,3}$,
two $V_i$'s and one $\bar V_i$ should get VEVs.
We require that
$\lambda_b\sim\lambda_{\tau}$ at the unification scale, which
imposes $\l \Phi_1^+\r\sim\l \Phi_1^- \r$.
We would like to stress that for different standard--like models,
requiring realistic quark mass matrices imposes similar
constraints. A solution that satisfies these requirements is
given by the following set of non--vanishing VEVs:
$$\{\Phi_1^\pm,\bar \Phi_1^\pm,\Phi_2^-,\bar \Phi_2^\pm,\Phi_3^\pm,
\bar \Phi_3^\pm,\Phi_{45},
\Phi_{13},\bar \Phi_{13},{\bar\Phi}_{23},
V_1,\bar V_2,V_3,\xi_1,\xi_2\} \eqno(28)$$
with
$$\eqalignno{
&-\l\Phi_3^-\r=\l\Phi_1^-\r=\l{\bar\Phi}_1^-\r=\l{\bar\Phi}_3^-\r=
  {3\over\sqrt{10}}{g\over{4\pi}}{1\over\sqrt{2\alpha^\prime}}&(29a)\cr
&\vert\l\Phi_1^+\r\vert^2=2\vert\l{\bar\Phi}_1^+\r\vert^2=
\vert\l\Phi_3^+\r\vert^2=\vert\l{\bar\Phi}_3^+\r\vert^2=
  {g^2\over{16\pi^2}}{1\over{2\alpha^\prime}}&(29b)\cr
&-\l\Phi_2^-\r=\l{\bar\Phi}_2^-\r=
  \left({\sqrt{2}-1}\over{\sqrt{2}}\right)^{1\over2}
  {g\over{4\pi}} {1\over\sqrt{2\alpha^\prime}}&(29c)\cr
&{1\over6}\vert\l\Phi_{45}\r\vert^2=
  {2}\vert\l{\bar\Phi}_{23}\r\vert^2=
  2\vert\l{\bar\Phi}_2^+\r\vert^2=
  {g^2\over{16\pi^2}}{1\over{2\alpha^\prime}}&(29d)\cr
&\l\Phi_{13}\r=  \left({\sqrt{2}-1}\over{\sqrt{2}}\right)^{1\over2}
  \left(1-{3\over\sqrt{5}}\left({\sqrt{2}-1}\over{\sqrt{2}}\right)^{1\over2}
  \right){g\over{4\pi}}{1\over\sqrt{2\alpha^\prime}}&(29e)\cr
&\vert\l{\bar\Phi}_{13}\r\vert^2=\vert\l\Phi_{13}\r\vert^2+
  {3\over2}{g^2\over{16\pi^2}}{1\over{2\alpha^\prime}}&(29f)\cr
&\vert\l{V_1}\r\vert^2={1\over3}\vert\l{\bar V}_2\r\vert^2=
{1\over2}\vert\l{V_3}\r\vert^2={g^2\over{16\pi^2}}{1\over{2\alpha^\prime}}
&(29g)\cr}$$
The VEVs of $\xi_1$ and $\xi_2$ are not constrained. With this set of VEVs, the
up and down quark mass matrices, $M_u$ and $M_d$ are given by
$$M_u\sim\left(\matrix{&0
&{{V_3{\bar V}_2\Phi_{45}\bar \Phi_3^+}\over{M^4}} &0\cr
&{{V_3{\bar V}_2\Phi_{45}\bar \Phi_2^+}\over{M^4}}
&{{{\bar\Phi}_i^-\bar \Phi_i^+}\over{M^2}}
&{V_1{\bar V}_2\Phi_{45}\bar \Phi_2^+}\over{M^4} \cr
&0 &{V_1{\bar V}_2\Phi_{45}{\bar\Phi}_1^+}\over{M^4}
&1\cr}\right)v_1\eqno(30)$$
and
$$M_d\sim\left(\matrix{&0
&{{V_3{\bar V}_2\Phi_{45}}\over{M^3}} &0\cr
&{{V_3{\bar V}_2\Phi_{45}\xi_1}\over{M^4}}
&{{{\bar\Phi}_2^-\xi_1}\over{M^2}} &{V_1{\bar V}_2\Phi_{45}\xi_i}\over{M^4} \cr
&0 &{V_1{\bar V}_2\Phi_{45}\xi_i}\over{M^4}
&{{\Phi_1^+\xi_2}\over{M^2}}\cr}\right)v_2\eqno(31)$$
with $v_1$, $v_2$ and $M$ as before.
The up and down quark mass matrices are diagonalized
by bi--unitary transformations
$$\eqalignno{&U_LM_uU_R^\dagger=D_u\equiv{\rm diag}(m_u,m_c,m_t),&(32a)\cr
	     &D_LM_dD_R^\dagger=D_d\equiv{\rm diag}(m_d,m_s,m_b),&(32b)\cr}$$
with the CKM mixing matrix given by
$$V=U_LD_L^\dagger.\eqno(33)$$
The VEVs of $\xi_1$ and $\xi_2$ are fixed to be $\langle \xi_1 \rangle
\sim M/12$ and $\langle \xi_2 \rangle \sim M/4$ by the masses $m_s$ and $m_b$
respectively. Substituting the
VEVs and diagonalizing $M_u$ and $M_d$ by a bi--unitary transformation, we
obtain the mixing matrix
$$\vert V \vert\sim \left(\matrix {0.98&0.205&0.002 \cr
                                  0.205&0.98&0.012 \cr
                                 0.0004&0.012&0.99 \cr} \right) \eqno(34)$$
To study the effect of the renormalization from the unification scale to the
electroweak scale we run the coupled renormalization group equations
of the MSSM in matrix form [\BKS]. The renormalization
does not affect the mixing terms that correspond to the
Cabibbo $2\times2$ submatrix by much. The remaining elements, that mix the
heavy generation with the lighter two generations are modified by up to thirty
percent. Therefore,
$V$ is a CKM matrix with elements of the correct order of magnitude. The
string model does not determine the flat direction (scalar VEVs)
and therefore does not predict
the matrix elements.
Since our aim is only to demonstrate the possibility of obtaining a
realistic CKM matrix and not to predict it, we do not pursue this point
further.

In Eq. (34) only the magnitude of the CKM matrix elements appear since we
took all the VEVs to be real. From Eq. (29) we see that the phases of the VEVs
except for those of $\Phi_1^-, \bar \Phi_1^-, \Phi_2^-, \bar
\Phi_2^-,\Phi_3^-, \bar \Phi_3^-, \Phi_{13}$ are not fixed by the F and D
constraints. By giving phases to some of these VEVs
we will be able to obtain a complex CKM matrix.

Consider the set of VEVs given in Eq. (29) where now we give phases to the VEVs
of $V_1$,$\bar V_2$ and $V_3$ only:
$$\eqalignno{&\l V_1 \r=e^{i\alpha}{g\over{4\pi}}
{1\over\sqrt{2\alpha^\prime}} &(35a) \cr
&\l \bar V_2 \r=e^{-i\beta}{\sqrt{3}}{g\over 4\pi}{\sqrt {1\over
2\alpha^\prime}} &(35b) \cr
&\l V_3 \r=e^{i\gamma}{\sqrt 2}{g\over {4\pi}}{\sqrt {1\over 2\alpha^
\prime}} &(35c) \cr}$$
where $\alpha$, $\beta$ and $\gamma$ are completely arbitrary. For this choice
of VEVs the mass matrices $M_u$ and $M_d$ become complex. Only the two
combinations $\alpha-\beta$ and $\gamma-\beta$ of these three
angles appear in $M_u$ and $M_d$. As a result we can always set one of the
angles to zero without loss of generality.
For the choice $tan(\gamma-\beta)=1$ and $tan(\alpha-\beta)={1}$ and using
the freedom to change the phases of 5 quarks we obtain
$$ V \sim \left(\matrix {0.973&0.230&0.002e^{-i{{19}\over{20}}\pi} \cr
                0.230&0.973&0.01 \cr
                0.0004e^{i{4\over5}\pi}&0.01&0.99 \cr} \right) \eqno(36)$$
This form of the CKM mixing matrix resembles the mixing matrix
in the Chau--Keung parametrization [\HL]. The 11, 12, 21, 23 and 33 matrix
elements are real (This can always be done by the phase transformations on the
5 quarks.) and
the 22 and 32 elements have small phases that we have neglected.
The parametrization--invariant CP violating quantity,
$\vert{J}\vert=\vert\hbox{Im}(V_{ij}V_{lk}V^*_{ik}V^*_{lj})\vert$ for
any $i\ne l$, $j\ne k$, is of the order of $10^{-6}$.
Experimentally, in the Standard Model, $\delta$ is $20^o<\delta<178^o$ [\NIR].
There are different possible choices of the phases $\alpha$,
$\beta$ and $\gamma$ which give different CP violating phases $\delta$.
In general, as $\alpha$, $\beta$ and $\gamma$ are varied continuously,
one obtains a mixing matrix with varying phase $\delta$ and non--negligible
phases also appearing in the 22 and 32 elements.
We do not discuss further the dependence of $\delta$ on the flat directions
as our aim is only to demonstrate the possibility of having a realistic
phase in the string model.

\bigskip
\centerline {\bf 6. Ansatze for mass matrices}

The standard model uses ten free parameters to parametrize the quark masses
and mixing. Several ansatze for the quark mass matrices have been proposed
to reduce the number of free parameters. These ansatze assume the existence
of discrete symmetries that force some of the entries in the
quark mass matrices to vanish. The origin of these ansatze and of the
symmetries that they assume to have is not explained.
In this section, we discuss the relation between the
quark mass matrices in the superstring standard--like models and a few
of these ansatze.

Consider for example the Fritzsch ansatz [\F] with

$$ M_u =\left(\matrix {0&a_u&0 \cr
                       a_u&0&b_u \cr
                       0&b_u&c_u \cr} \right){\hskip 1cm}
   M_d =\left(\matrix {0&a_d&0 \cr
                       a_d^*&0&b_d \cr
                       0&b_d^*&c_d \cr} \right) \eqno(37a,b)$$
with all elements of $M_u$ and $c_d$ real. From the mixing terms given by Eqs.
(16--20) we see that a mass matrix of the form of Eq. (37b)
cannot be obtained in our model.
The reason is that in Eq. (37b) the elements
$23$ and $32$ in $M_d$ are complex conjugates of each other, while in model 1,
they (and therefore their phases) are equal. However, we can
consider modifying Eq. (37b) by taking
$$M_{\{u,d\}} =\left(\matrix {0&a&0 \cr
                       a&0&b \cr
                       0&b&c \cr} \right) \eqno(38)$$
with $a$, $b$, $c$ complex. We see that this
ansatz can be obtained from our model
for a flat direction with
$\l \bar \Phi_i^+ \bar \Phi_i^-\r =0$ and $\l \bar \Phi_2^-\r =0$
and non--vanishing VEVs for $V_1$,
$\bar V_2$ and $V_3$. Similarly, consider an ansatze of the form
$$ M_{\{u,d\}} =\left(\matrix {0&0&a \cr
                       0&0&b \cr
                       a&b&c \cr} \right) \eqno(39a)$$
This form of mass matrices can be obtained in our model
by a flat direction with the same conditions but VEVs for ${\bar V}_1$,
$V_2$ and $V_3$. This form of mass matrices is not realistic as it gives
vanishing up and down quark masses and zero Cabibbo angle. However,
Eq. (39) illustrates the dependence of the vanishing mass matrix
entries on the state from the sectors $b_j+2\gamma$ that obtain
non--vanishing VEVs. Similarly, an ansatze of the form
$$ M_{\{u,d\}} =\left(\matrix {0&a&b \cr
                       a&0&0 \cr
                       b&0&c \cr} \right) \eqno(39b)$$
can be obtained with non--vanishing VEVs for $V_1$, $V_2$ and $\bar V_3$.

The superstring standard--like models provide the possibility
of obtaining up and down mass matrices of different textures.
Non--vanishing VEVs for $V_1$, $\bar V_2$, $V_3$, $\xi_1$ and $\xi_2$
produce down quark mass that give a realistic CKM matrix. If we impose
a flat F and D solution with vanishing VEV for ${\bar\Phi}^+_2$ or
${\bar\Phi}^+_3$, the up quark mass matrix will take the form
$$ M_u =\left(\matrix {0&a&0 \cr
                       0&b&0 \cr
                       0&c&1 \cr} \right){\hskip .5cm}{\hbox{or}}
{\hskip .5cm} M_u =\left(\matrix {0&0&0 \cr
	                          a&b&c \cr
           	                  0&c&1 \cr} \right) \eqno(40)$$
This form of up mass matrix results in vanishing up quark mass being
the well known solution to the strong CP problem [\SCP]. The down quark
mass matrix is of the form of Eq. (38), and therefore $m_u=0$ while
$m_d\ne0$.
To ascertain if this solution to the strong CP problem
is a possible solution within the context of the
superstring standard--like models, one would have to check the possible
sources for the 12, 21 and 11 entries in the up quark mass matrix.
Those being nonrenormalizable terms up to a sufficient order and possible
contributions to the diagonal up quark mass term from VEVs that break
$U(1)_{Z^\prime}$ [\NRT].

The vanishing mixing entries in Eqs. (30--31) arise because only one $V_i$
from each
sector $b_i+2\gamma$ obtained a non--vanishing VEV. If more than one state
from any sector $b_i+2\gamma$ obtain a non--vanishing VEV,
for example $\{V_1,{\bar V}_2,V_3,{\bar V}_3\}$,
then in general all the off--diagonal terms will be nonzero at some order
of nonrenormalizable terms. However, one has to be careful not to generate
Higgs mass terms that will render all the Higgs doublets superheavy.
Thus, it is seen that the vanishing off--diagonal entries in the quark mass
matrices arise due to effective discrete symmetries which are a result of
the particular solution that we choose in our model. The mixing terms
Eqs. (17--20) show that for different F and D flat directions
different textures of mass matrices can be obtained. The resulting
mass matrices are not necessarily symmetric. A sufficient condition for
the down quark mass matrix to have off--diagonal terms is that two $V_i$
one ${\bar V}_i$, $\Phi_{45}$ and $\xi_i$ obtain non--vanishing VEVs.
The up quark mass matrix may be diagonal for some choices of flat
directions. The richness of the flat F and D flat solution space gives us the
reason to hope that the superstring standard--like models can
successfully reproduce the observed quark mixing and mass spectrum.

\bigskip
\centerline {\bf 7. Conclusions}

In this paper, we examined the three generation mixing among quark families in
realistic, superstring derived standard--like models.
Family mixing is induced by
order $N>3$ terms in the superpotential which respect all local and global
symmetries of the string model. From the explicit mixing terms we observe that
generation mixing arises from certain sectors of the spectrum, namely the
sectors $b_j+2\gamma$. The states in these sectors (which are in the vector
representations of $SU(5)_H \times SU(3)_H$) obtain VEVs by the F and D
flatness conditions which are essential to preserve SUSY
close to the Planck scale. By
examining the conservation of left--handed global $U(1)_\ell$ symmetries,
we showed
that the existence of the sectors $b_j+2\gamma$ is a necessary condition for
obtaining nonrenormalizable quark mixing terms. Since the
sectors $b_j+2\gamma$ always exist in the free fermionic
models with fermion generations from the
twisted sectors $b_j$, and with a $Z_4$ twist (vector $\gamma$), generation
mixing is a generic feature of these models. The question then is whether
there exists
a suitable F and D flat direction which gives a realistic CKM matrix.

There exist F and D flat directions that produce three generation mixing.
We found one such flat direction given by Eq. (29) and calculated the mixing
matrix from it. In addition, by giving phases
to some of the scalar VEVs, one can obtain the weak CP
violating phase, $\delta$, in the CKM matrix. In our model, we were able to
obtain $\delta$ by giving phases only to the VEVs of $V_1$, $\bar V_2$ and
$V_3$.
We emphasize that the string model does not fix the flat direction and
therefore
does not predict the CKM matrix. Our aim is only to show that a realistic CKM
matrix can be obtained in this class of models.

Can we improve our order of magnitude results? There are three ways that our
results can be made more predictive. First, we can take into account the
coefficients $c$ which enter the effective Yukawa couplings.
In this paper we have taken them to be of order $O(1)$.
They can, in principle, be calculated from the correlators of vertex operators
for every nonrenormalizable term.
Second, we can consider other phenomenological constraints on the
model such as acceptable neutrino masses, baryon decay, very small FCNCs etc.,
in addition to the ones we took into account. These will further constrain the
possible F and D flat directions and make the model more predictive. For
example, the condition of realistic neutrino masses requires the existence
of light $SO(10)$ singlet fermions which mix with the right--handed neutrinos,
$N_i$ [\FHn]. This condition puts additional
constraints on the possible F and D
flat directions which in turn constrain the possible CKM matrices in the model.
The weak CP violating
phase $\delta$ can always be obtained from the phases of $V_i$ and $\Phi_{45}$
which enter only the D flatness equations that are quadratic in VEVs.
Therefore,
their phases and those of $\xi_{1,2}$ (which do not appear in F and D flatness
equations) are completely free. In this paper, we
considered phases only for $V_i$ since this is the simplest possibility. A
better fix on $\delta$ can be obtained by giving phases to all the VEVs above.
The freedom in the flat directions gives us reason to believe that
realistic quark mixing and mass spectrum can be obtained from the superstring
standard--like models.

The horizontal symmetries that arise in superstring models due to the
compactification from ten to four dimensions constrain the allowed terms
in the superpotential and consequently the terms in the fermion mass matrices.
In this paper we examined the superstring derived standard--like models.
These models are constructed in the free fermionic formulation and
correspond to models that are based on $Z_2\times Z_2$ orbifold
compactification. We showed that the horizontal symmetries and the choice
of flat directions in the application of the Dine--Seiberg--Witten
mechanism constrain some of the entries in the quark mass matrices
to vanish. Consequently different textures for the fermion mass matrices
may be obtained from the standard--like models, that may naturally resolve
some of the problems that exist in traditional GUTs. For example, the
relation $\lambda_t=\lambda_b$ in $SO(10)$ models that forces a large
value for $\tan\beta={{v_1}/{v_2}}$, is broken in the superstring
standard--like models and allows small values for $\tan\beta$. Similarly,
the choice of flat directions may naturally lead to $m_u=0$ with $m_d\ne0$.
Thus, the superstring standard--like models may provide simple
and well motivated solutions to some of the fundamental problems in
particle physics. We will expand upon the phenomenology derived from these
models in future publications.

\bigskip
\centerline{\bf Acknowledgments}

This work is supported in part by the Feinberg School.
We would like to thank Miriam Leurer and Yossi Nir for valuable discussions.
Edi Halyo would like to express his gratitude to Isak and Gulen Halyo
for their support during this work.

\refout

\vfill
\eject

\end

\input tables.tex
\nopagenumbers
\tolerance=1200

{\hfill
{\begintable
\  \ \|\ ${\psi^\mu}$ \ \|\ $\{{\chi^{12};\chi^{34};\chi^{56}}\}$\ \|\
{${\bar\psi}^1$, ${\bar\psi}^2$, ${\bar\psi}^3$,
${\bar\psi}^4$, ${\bar\psi}^5$, ${\bar\eta}^1$,
${\bar\eta}^2$, ${\bar\eta}^3$} \ \|\
{${\bar\phi}^1$, ${\bar\phi}^2$, ${\bar\phi}^3$, ${\bar\phi}^4$,
${\bar\phi}^5$, ${\bar\phi}^6$, ${\bar\phi}^7$, ${\bar\phi}^8$} \crthick
$\alpha$
\|\ 0 \|
$\{0,~0,~0\}$ \|
1, ~~1, ~~1, ~~0, ~~0, ~~0 ,~~0, ~~0 \|
1, ~~1, ~~1, ~~1, ~~0, ~~0, ~~0, ~~0 \nr
$\beta$
\|\ 0 \| $\{0,~0,~0\}$ \|
1, ~~1, ~~1, ~~0, ~~0, ~~0, ~~0, ~~0 \|
1, ~~1, ~~1, ~~1, ~~0, ~~0, ~~0, ~~0 \nr
$\gamma$
\|\ 0 \|
$\{0,~0,~0\}$ \|
{}~~$1\over2$, ~~$1\over2$, ~~$1\over2$, ~~$1\over2$,
{}~~$1\over2$, ~~$1\over2$, ~~$1\over2$, ~~$1\over2$ \| $1\over2$, ~~0, ~~1,
{}~~1,
{}~~$1\over2$,
{}~~$1\over2$, ~~$1\over2$, ~~0 \endtable}
\hfill}
\smallskip
{\hfill
{\begintable
\  \ \|\
${y^3y^6}$,  ${y^4{\bar y}^4}$, ${y^5{\bar y}^5}$,
${{\bar y}^3{\bar y}^6}$
\ \|\ ${y^1\omega^6}$,  ${y^2{\bar y}^2}$,
${\omega^5{\bar\omega}^5}$,
${{\bar y}^1{\bar\omega}^6}$
\ \|\ ${\omega^1{\omega}^3}$,  ${\omega^2{\bar\omega}^2}$,
${\omega^4{\bar\omega}^4}$,  ${{\bar\omega}^1{\bar\omega}^3}$  \crthick
$\alpha$ \|
1, ~~~0, ~~~~0, ~~~~0 \|
0, ~~~0, ~~~~1, ~~~~1 \|
0, ~~~0, ~~~~1, ~~~~1 \nr
$\beta$ \|
0, ~~~0, ~~~~1, ~~~~1 \|
1, ~~~0, ~~~~0, ~~~~0 \|
0, ~~~1, ~~~~0, ~~~~1 \nr
$\gamma$ \|
0, ~~~1, ~~~~0, ~~~~1 \|\
0, ~~~1, ~~~~0, ~~~~1 \|
1, ~~~0, ~~~~0, ~~~~0  \endtable}
\hfill}
\smallskip
\parindent=0pt
\hangindent=39pt\hangafter=1
\baselineskip=18pt

{{\it Table 1.} A three generations ${SU(3)\times SU(2)\times U(1)^2}$
model. The choice of generalized GSO coefficients is:
${c\left(\matrix{b_j\cr
                                    \alpha,\beta,\gamma\cr}\right)=
-c\left(\matrix{\alpha\cr
                                    1\cr}\right)=
c\left(\matrix{\alpha\cr
                                    \beta\cr}\right)=
-c\left(\matrix{\beta\cr
                                    1\cr}\right)=
c\left(\matrix{\gamma\cr
                                    1,\alpha\cr}\right)=
-c\left(\matrix{\gamma\cr
                                    \beta\cr}\right)=
-1}$ (j=1,2,3),
with the others specified by modular invariance and space--time
supersymmetry.
Trilevel Yukawa couplings are obtained only for
${+{2\over3}}$ charged quarks.
The $16$ right--moving
internal fermionic states
$\{{\bar\psi}^{1,\cdots,5},{\bar\eta}^1,
{\bar\eta}^2,{\bar\eta}^3,{\bar\phi}^{1,\cdots,8}\}$,
correspond to the $16$ dimensional compactified  torus of the ten dimensional
heterotic string.
The 12 left--moving and 12 right--moving real internal fermionic states
correspond to the six left
and six right compactified dimensions in the bosonic language.
$\psi^\mu$ are the two space--time
external fermions in the light--cone gauge and
$\chi^{12}$, $\chi^{34}$, $\chi^{56}$
correspond to the spin connection in the bosonic constructions.}
\vskip 2.5cm

\vfill
\eject

\input tables.tex
\baselineskip=18pt
\hbox
{\hfill
{\begintable
\ F \ \|\ SEC \ \|\ $SU(3)_C$ $\times$ $SU(2)_L$ \ \|\
$Q_C$ & $Q_L$ & $Q_1$ & $Q_2$
 & $Q_3$ & $Q_4$  & $Q_5$ & $Q_6$
 \ \|\ $SU(5)$ $\times$ $SU(3)$ \ \|\ $Q_7$ &
$Q_8$  \crthick
$V_1$ \|\ ${b_1+2\gamma}+(I)$ \|(1,1)\|~0 & ~~0 & ~~0 & ~~${1\over 2}$ &
 ~~$1\over 2$ & ~~$1\over2$
 & ~~0 & ~~0 \|(1,3)\| $-{1\over 2}$ &
{}~~$5\over 2$   \nr
${\bar V}_1$ \|\                \|(1,1)\| ~~0 & ~~0 & ~~0 & ~~${1\over 2}$ &
 ~~$1\over 2$ & ~~$1\over2$
  & ~~0 & ~~0
  \|(1,$\bar 3$)\| ~~${1\over 2}$ &
$-{5\over 2}$  \nr
$T_1$ \|\                \|(1,1)\| ~~0 & ~~0 & ~~0 & ~~${1\over 2}$ &
 ~~$1\over 2$ & $-{1\over2}$
  & ~~0 & ~~0
  \|(5,1)\| $-{1\over 2}$ &
$-{3\over 2}$  \nr
${\bar T}_1$ \|\                \|(1,1)\| ~~0 & ~~0 & ~~0 & ~~${1\over 2}$ &
 ~~$1\over 2$ & $-{1\over2}$
   & ~~0 & ~~0
  \|($\bar 5$,1)\| ~~$1\over2$ &
{}~~${3\over 2}$  \cr
$V_{2}$ \|\ ${b_2+2\gamma}+(I)$ \|(1,1)\| ~~0 & ~~0 &
 ~~${1\over 2}$ & ~~0 &
 ~~$1\over 2$ & ~~0 &  ~~$1\over2$ & ~~0
 \|(1,3)\| $-{1\over 2}$ &
 ~~$5\over 2$  \nr
${\bar V}_{2}$ \|\                \|(1,1)\| ~~0 & ~~0 & ~~${1\over 2}$ & ~~0 &
 ~~$1\over 2$ & ~~0 & ~~$1\over2$ & ~~0
  \|(1,$\bar 3$)\| ~~${1\over 2}$ &
 $-{5\over 2}$  \nr
$T_{2}$ \|\                \|(1,1)\| ~~0 & ~~0 & ~~${1\over 2}$ & ~~0 &
 ~~$1\over 2$ & ~~0 & $-{1\over2}$ & ~~0
  \|(5,1)\| $-{1\over 2}$ &
 $-{3\over 2}$ \nr
${\bar T}_{2}$ \|\                \|(1,1)\| ~~0 & ~~0 & ~~${1\over 2}$ & ~~0 &
 ~~$1\over 2$ & ~~0 & $-{1\over2}$ & ~~0
  \|($\bar 5$,1)\| ~~$1\over 2$ &
 ~~${3\over 2}$  \cr
$V_{3}$ \|\ ${b_3+2\gamma}+(I)$ \|(1,1)\| ~~0 & ~~0 & ~~${1\over 2}$ &
 ~~${1\over 2}$ & ~~0 & ~~0 & ~~0 & ~~${1\over2}$
       \|(1,3)\| $-{1\over 2}$ &
 ~~$5\over 2$  \nr
${\bar V}_{3}$ \|\                \|(1,1)\| ~~0 & ~~0 & ~~${1\over 2}$ &
 ~~${1\over 2}$ & ~~0 & ~~0 & ~~0 & ~~$1\over2$
 \|(1,$\bar 3$)\| ~~${1\over 2}$
 & $-{5\over 2}$  \nr
$T_{3}$ \|\                \|(1,1)\| ~~0 & ~~0 & ~~${1\over 2}$ &
 ~~${1\over 2}$ & ~~0 & ~~0 & ~~0 & $-{1\over2}$
       \|(5,1)\|
 $-{1\over 2}$ & $-{3\over 2}$  \nr
${\bar T}_{3}$ \|\                \|(1,1)\| ~~0 & ~~0 & ~~${1\over2}$ &
 ~~${1\over 2}$ & ~~0 & ~~0 & ~~0 & $-{1\over2}$
   \|($\bar 5$,1)\| ~~$1\over 2$
 & ~~${3\over 2}$
 \endtable}
\hfill}
\bigskip
\parindent=0pt
\hangindent=39pt\hangafter=1
{\it Table 2.} Massless states from the sectors $b_j+2\gamma$,
and their quantum numbers.

\vfill
\eject

\end
\bye